# Dynamical impact of the Mekong River plume in the South China Sea

Xiyuan Zeng[1,2], Annalisa Bracco[1*], and Filippos Tagklis[1,3]

[1]School of Earth and Atmospheric Sciences, Georgia Institute of Technology, Atlanta, GA, USA

[2]Program in Ocean Sciences and Engineering, Georgia Institute of Technology, Atlanta, GA, USA

[3] The NOAA Cooperative Institute for Marine And Atmospheric Studies (CIMAS) at the University of Miami's Rosenstiel School of Marine and Atmospheric Science, Miami, FL, USA

Corresponding author: Annalisa Bracco (abracco@gatech.edu)

**Key Points: South China Sea; Regional Ocean Modeling; Freshwater induced dynamics**

**Abstract**

Near the ocean surface, river plumes influence stratification, buoyancy and transport of tracers, nutrients and pollutants. The extent to which river plumes influence the overall circulation, however, is generally poorly constrained. This work focuses on the South China Sea (SCS) and quantifies the dynamical impacts of the Mekong River plume, which is bound to significantly change in strength and seasonality in the next 20 years if the construction of over hundred dams moves ahead as planned. The dynamic impact of the freshwater fluxes on the SCS circulation are quantified by comparing submesoscale permitting and mesoscale resolving simulations with and without riverine input between 2011 and 2016. In the summer and early fall, when the Mekong discharge is at its peak, the greater stratification causes a residual mesoscale circulation through enhanced baroclinic instability. The residual circulation is shaped as an eddy train of positive and negative vorticity. Submesoscale fronts are responsible for transporting the freshwater offshore, shifting eastward the development of the residual mesoscale circulation, and further strengthening the residual eddy train in the submesoscale permitting case. Overall, the northward transport near the surface is intensified in the presence of riverine input. The significance of the mesoscale-induced and submesoscale-induced transport associated with the river plume is especially important in in the second half of the summer monsoon season, when primary productivity has a secondary maximum. Circulation changes, and therefore productivity changes, should be anticipated if human activities modify the intensity and seasonality of the Mekong River plume.

**Plain Language Summary**

We explore how the freshwater forcing induced by the Mekong River plume impact the ocean circulation in the South China Sea with a suite of regional ocean simulations. We focus on the dynamics associated to the river plume in boreal summer. The motivation for this investigation is the planned construction of more than hundred hydropower dams along the Mekong River Basin and its major tributaries. The dams will dramatically reduce the Mekong River annual mean flow and its seasonal cycle (up to an order of magnitude) and sediment loading. Our results suggest that the overall circulation of the basin may be impacted by these changes, through changes in the mesoscale circulation and instability characteristics of the flow.

We argue that a reduced productivity of the offshore water of the South China Sea along the pathway of the summer jet may be an undesirable outcome as well.

**1 Introduction**

The South China Sea (SCS) is the largest marginal sea in Southeast Asia. It occupies the region from the equator to 23°N and from 99°E to 121°E with a maximum depth of over 5000 m. Two broad shelves with depth shallower than 200 m are located on the northwestern and southwestern SCS, bordering the central, bowl-shaped deep basin (Hu et al. 2000). The SCS circulation is strongly influenced by the monsoonal winds (Hellerman et al., 1983). In winter, northeasterly winds prevail over the whole region with an average magnitude of 9 ms$^{-1}$ and the coastal upwelling off the coast of Vietnam is suppressed due to the wind-driven cyclonic circulation cell (Chao et al. 1996). In contrast, the weaker southwest summer monsoon winds from June to September, with an average magnitude of 6 ms$^{-1}$, induce coastal upwelling that brings colder and saltier water to the surface (Wang et al., 2015). As a result of the wind upwelling, two distinct circulation cells are visible during this season (Dippner et al. 2011). The summer monsoon also brings intense rainfall over the ocean and Indochina, which drives large pulses of riverine flow into the SCS via the Mekong River (Shaw and Chao 1994). The Mekong River originates in the Tibetan Plateau in China, drains a nearly 800,000 km$^2$ watershed and discharges into the SCS at the Mekong Delta in Vietnam with an annual mean flow of ~15,000 km$^3$ year$^{-1}$ (Pokhrel et al. 2018). The river plume contributes significantly to the hydrodynamics near and offshore its outflow points (Gonzalez et al. 2018), as a fresh offshore jet forms at about 12°N in late spring. The river water is then advected to the northeast by the monsoonal winds throughout the summer. The jet results from the blockage of the southwest monsoon by the mountain range on the east coast of Indochina and is characterized by a strong intraseasonal variability in strength and extension (Xie et al., 2007). The path followed by the Mekong outflow differs between seasons: the river plume remains confined in the southward coastal jet during winter and spreads northwards over the shelf and then off-shore, transported by the jet, in summer. The circulation in the basin is characterized by strong horizontal mixing and sustained mesoscale activity (Gan et al. 2006, Cardona & Bracco 2012). The eddy field in summer and fall modulates the propagation of the river plume water and is especially energetic between 7°N and 15°N (Zhuang et al., 2020). The interplay between coastal upwelling, the jet, riverine (and nutrients) inflow and mesoscale circulations modulates phytoplankton communities and in turn primary productivity and fish abundance in the region (Weber et al., 2019), and controls the secondary peak in primary productivity observed between August and October (Liu and Chai, 2009).

Here, we explore the role of freshwater forcing on the SCS circulation using a regional ocean model. We focus on the dynamics associated to the river plume in boreal summer, and we explore its interaction with the jet and the eddy train observed in this season (Nan et al., 2011; Chen et al., 2011; Zheng et al., 2014). The motivation for this work is the planned construction of 123 hydropower dams, including eleven hydropower plants, along the Mekong River Basin and its major tributaries. The dams are expected to dramatically reduce the Mekong River mean, seasonal flow cycle (up to an order of magnitude) and sediment loading (Wild and Loucks, 2015). Studies on the environmental impacts have been limited so far to the land ecosystems and water resources (e.g. Yoshida et al., 2020 for an up-to-date analysis), but our results suggest that the overall circulation of the basin may be impacted as well.

## 2. The Circulation in the SCS in Boreal Summer and Fall

The South China Sea is the largest marginal sea in South Asia. It connects to the Sulu and Java Seas in the south through several shallow passages, and with the Pacific Ocean through the deep (~ 2000 m) Luzon Strait in the north. Exchange of shelf waters between the South China Sea and the East China Sea occurs through the Taiwan Strait at a sill depth of 60 m.

As mentioned, the basin exhibits a strong, seasonally varying, wind-driven circulation characterized by coastal upwelling and downwelling associated to monsoonal winds (Wyrtki et al., 1961). The southwest summer monsoon drives the circulation between May and September (Hellerman et al., 1983). Monsoonal winds make their appearance in the central portion of the basin in May and expand over the entire basin in July and August. In September, winds reverse to the northeast or winter monsoon phase north of 20°N, and the northeast monsoon expands southward in October, covering the entire South China Sea by December. April marks the end of the winter monsoon season.

In response to the changing winds, a current develops off the coast of Vietnam. In Wyrtki's charts, the current flows northeastward in summer and the overall circulation pattern is an anticyclonic gyre. An important feature, significant along the northern boundary of the basin, is the Kuroshio intrusion through the Luzon Strait. Among the three passages along the eastern boundary, the Luzon Strait is the only opening where there is significant water exchange between the South China Sea and other basins. Water originating in the Pacific at depths between 1500 and 2000 m enters the SCS through the Luzon Strait and forms the deepest water mass of the basin (Nitani et al., 1972), moving southward into the basin along the western boundary (Shaw and Chao, 1994). In summer, two anticyclonic eddies appear to the southeast of the Zhongnan Peninsula and between the Xisha Islands and the Zhongsha Islands (Xu et al., 1982).

Meanwhile, a cyclonic eddy, induced by the local summertime upwelling, forms to the east of the Vietnam coast and forces the main circulation to meander. Wu et al. (1998) suggested that two modes in the velocity field could describe the seasonal and inter-annual variations of the SCS circulation. The first mode, associated with a southern gyre, shows symmetric seasonal reversal, being cyclonic in winter and anticyclonic in summer with limited year-to-year variations. The second mode, which contributes to the northern gyre off Vietnam, is responsible for the asymmetric seasonal and inter-annual variations. In summer, the north cyclonic gyre and the anticyclonic southern gyre form a system with a jet among them that separates from the coast of Vietnam at 11°N ~ 14°N. As mentioned, the jet transport large pulses of Mekong River water offshore.

The amount of river discharge varies in response to the monsoonal rainfall. Generally, the Mekong River outflow remains low from February to June, increases from July onward, reaches its maximum around October, and decreases sharply afterward. The Mekong outflow undergoes significant interannual variability. Over the period considered here, the Mekong outflow was strong in 2011, and weaker in 2015 and 2016 due to El Niño conditions. The El Niño Southern Oscillation (ENSO) indeed modulates the monsoonal winds, surface temperatures and precipitation over the SCS. It is associated to lower (higher) than normal summer rainfall following a positive or El Niño (negative or La Niña) phase in winter (Fan et al. 2018).

**3 Model Setup, Domain and Forcing**

**3.1 Model setup, domain and forcing**

We adopt the Regional Oceanic Modeling System-Coastal and Regional Ocean COmmunity model (ROMS-CROCO), a free-surface, terrain-following, hydrostatic, primitive-equation model (Marchesiello et al. 2009, Debreu et al., 2012). The model domain (Figure 1) extends to the whole SCS at 5 km horizontal resolution (LR for Low Resolution) resolving the mesoscale circulation with a nested area between 100°E-120°E and 3°N – 18°N at 1.6 km horizontal resolution (HR for Higher Resolution) that permits submesoscale circulations (McWilliams, 2016) to form. In the vertical there are 60 terrain-following layers with surface enhancement (no less than 20 layers in the upper 200 m in the deepest areas). After an opportune spin-up period in the LR configuration, CROCO is run from December 2009 to December 2016. The first year is discarded in the nested run and we focus on December 2010 onward in this work.

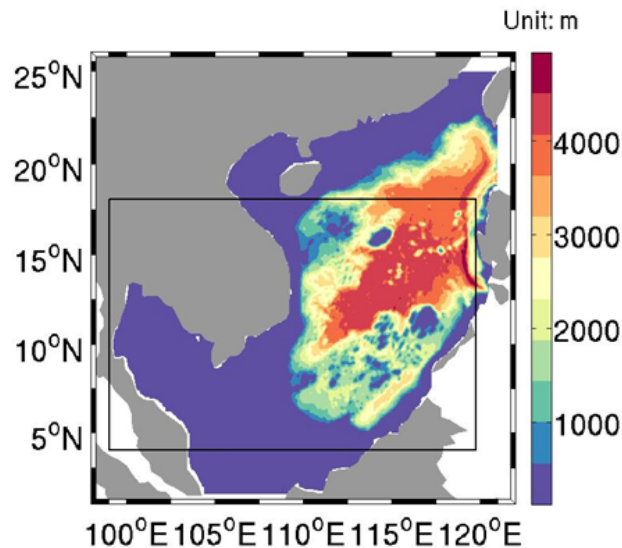

**Figure 1** Model Domain and bathymetry. The whole map highlights the domain of the 5 km (LR) case. The black box encompasses the area where CROCO is run at horizontal resolution of 1.6 km (HR).

The model bathymetry is interpolated from ETOPO2 (Sandwell and Smith 1997) and smoothed with a maximum allowed slope of 0.3. The parent grid simulation is forced by 6-hour surface wind stresses and daily heat fluxes from the European Centre for Medium-Range Weather Forecast ERA-5 reanalysis (Hersbach et al. 2020). Initial and boundary conditions are extracted from the Simple Ocean Data Assimilation (SODA), Version 3.4.2 dataset (Carton et al., 2018). For the nested simulations with 1.6 km horizontal resolution initial and boundary conditions are interpolated from the 5 km horizontal resolution case with identical heat and momentum forcing. To explore the impact of the Mekong plume on the circulation of this basin, we perform simulations with and without riverine input at both resolutions. With and without river cases for the 5 km and 1.6 km horizontal resolution run are indicated by LR-WR, LR-NoR, NoR and WR, respectively, here after, and we focus our analysis on the NoR and WR simulations.

Whenever the riverine forcing is included, the daily freshwater discharge from the stations of CanTho and Mythuan (Figure 2) is converted to an equivalent surface freshwater flux that decays exponentially away from the river mouths with a constant rate at each resolution (see Barkan et al., 2017 for a similar implementation strategy in the Gulf of Mexico). To isolate the impact from the fresh signals, no further salinity nudging is performed. The NoR case has the southern salinity boundary condition modified to remove the strong western boundary fresh signal that originates from east of Malaysia due to the Mekong riverine water accumulating in the shallower portion of the Gulf of Thailand.

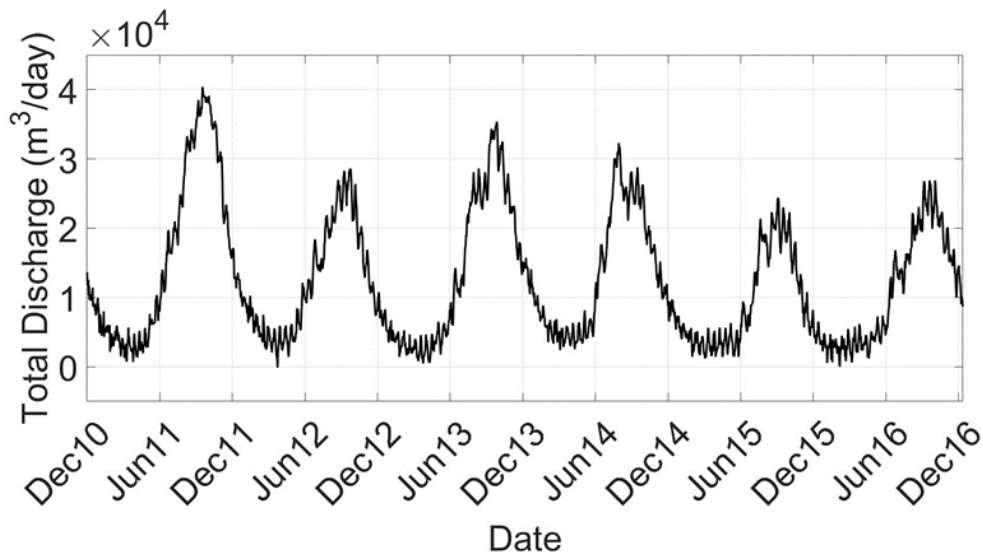

**Figure 2** Timeseries of the Mekong River outflows from year 2010 to 2016. Unit: $m^3$ $day^{-1}$.

### 3.2 Model validation

Mixed layer depth (MLD) is an important indicator of model realism given the processes we are interested in. We calculated the MLD using both a temperature criterion and a density criterion (Montégut et al., 2004). $MLD_T$ is often defined as the depth at which the temperature differs from that at the surface by 0.2 °C, while $MLD_\rho$ is commonly defined as the depth at which the density differs from the surface by 0.03 kg/m$^3$. The $MLD_T$ time series in Figure 4a are calculated for the whole domain and both WR and NoR integrations from 2011 to 2016. The MLD$_\rho$ annual cycle (not shown) shares similar trends with a small difference in the magnitude of the changes and larger differences among the two runs given the different riverine treatment. MLD$_T$ reaches its minimum in April, when the heat fluxes are strongest and the riverine discharge is at its minimum, and is deepest around December and January.

The model captures well the MLD$_T$ variability. A secondary peak around July that can be observed in 2012, 2013 and 2014 is also represented. Overall, the WR case has a slightly shallower MLD$_T$ compared to NoR. The difference is negligible in late winter to spring and more distinct during the months characterized by strong riverine discharge (Figure 3). The model underestimates the MLD in winter compared to SODA reanalysis, especially in 2013. We do not have in-situ data in the area of interest to verify if this is a bias in CROCO or results from a submesoscale-induced restratification of the water column in winter (Yu et al., 2019) that cannot be captured by SODA due to its coarser resolution.

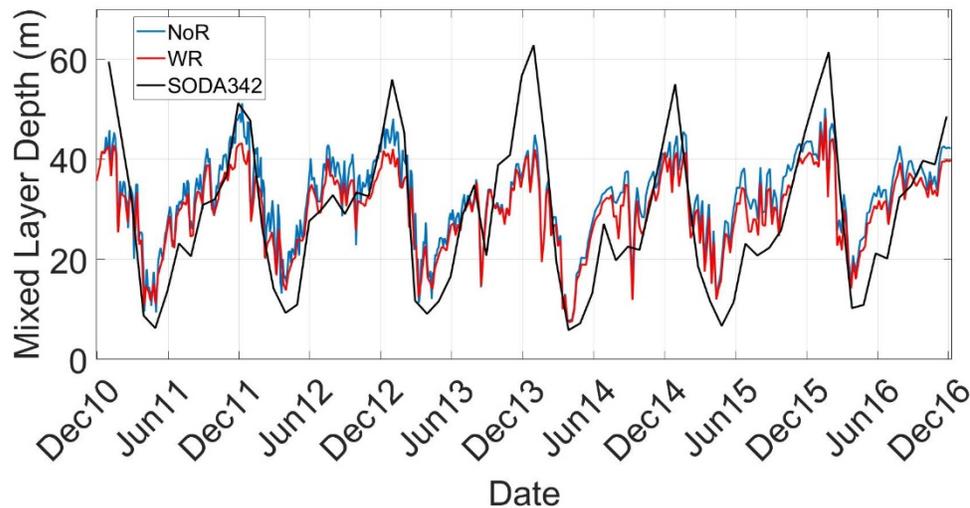

**Figure 3** Mixed layer depth time-series comparison among NoR, WR and SODA 3.4.2 calculated over the whole 1.6 km domain from the year 2011 to 2016 using a temperature criterion with δT= 0.2°C.

## 4 Modeled Circulation

As mentioned, the circulation along the Vietnam coast is impacted by the monsoonal winds and riverine discharge. The SCS western boundary current (SCSwbc) is a primary contributor for the redistribution of water properties, momentum, energy and nutrients. The current is mainly wind-driven (Liu et al., 2001) and influences the regional climate (e.g., Xie et al., 2003; Zhou et al., 2010).

In boreal summer, the SCSwbc flows from the Karimata Strait northward along the southern Malay Peninsula, passes by the Gulf of Thailand and flows along the southeastern Vietnamese coast. The monsoonal winds transport the river plume northeastward, forming a jet of freshwater, and excite strong coastal upwelling near the coastline of Vietnam from 9°N to 12°N. When the summer SCSwbc reaches approximately 11°N, it bends to the east, forming the so-called Southeast Vietnam Offshore Current (VOC, Fang et al., 2002). The exact position of the VOC separation varies among years between 10°N and 14°N (Figure 4).

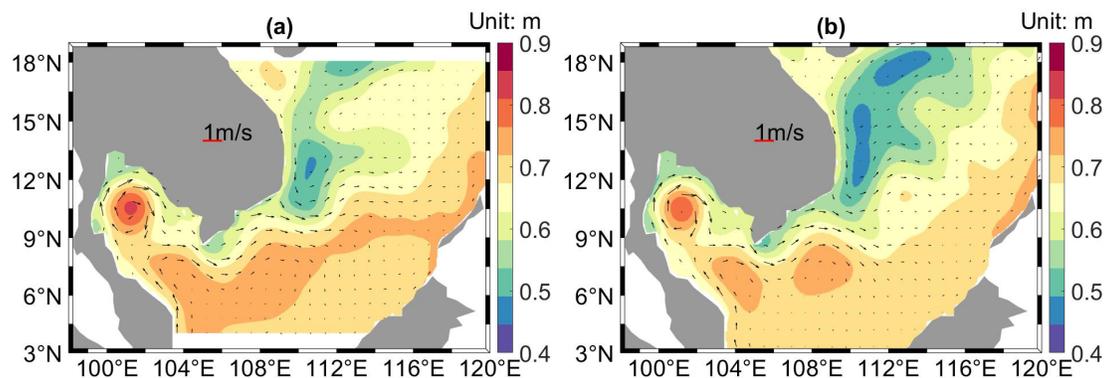

**Figure 4.** Sea surface high (SSH) field with superpose the surface velocity field averaged over June to October and 2011-2016 in the (a) WR and (b) LR-WR integrations.

The upwelling associated with the VOC leads to intense, negative temperature anomalies in the upper 30-40 m of the water column. Figures 5 and 6 show SSH anomalies, temperature and salinity snapshots at 15 meters depth for the WR and LR-WR runs in June and August 2012 as example.

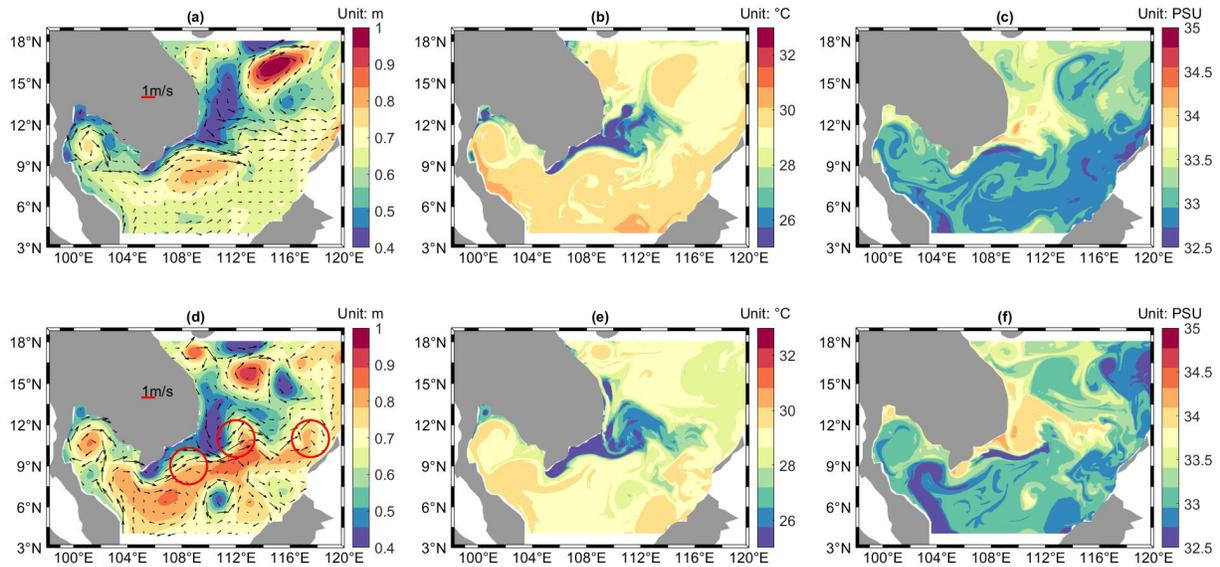

**Figure 5** SSH anomalies (left, a and d), temperature (center, b and e) and salinity (right, c and f) snapshots in June 15 and August 14, 2012 in the WR simulation. Temperature and salinity are shown at 15 m depth. The anticyclonic eddy train is indicated in the August snapshot.

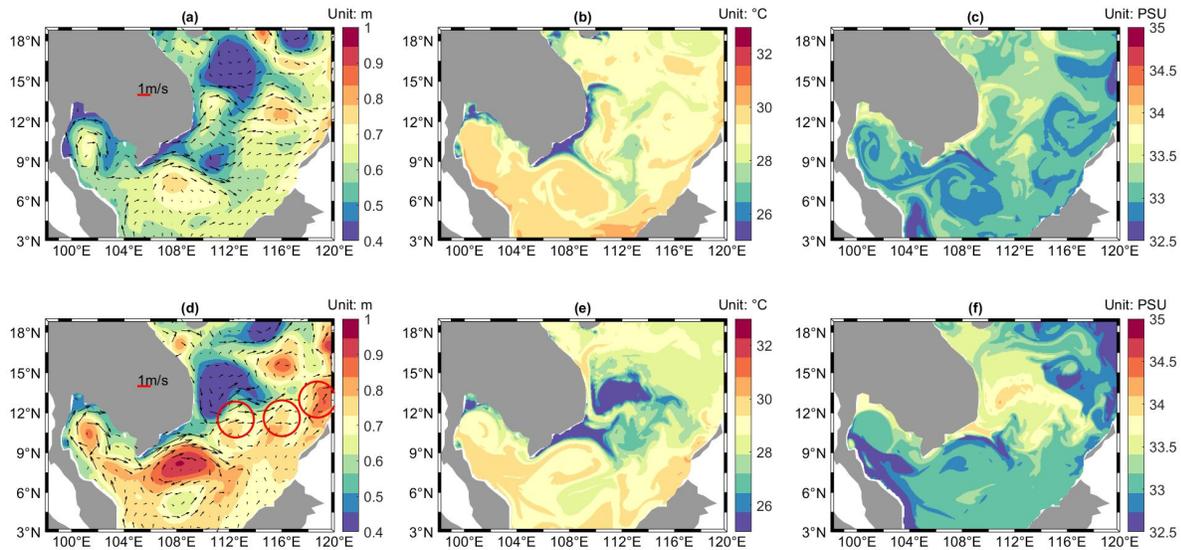

**Figure 6** As in Figure 5 but for the LR-WR simulation.

In June, the SW monsoon induces coastal upwelling, bringing colder and saltier water to the surface. The riverine signal penetrates eastward from June to August with interannual differences in timing and location. As the Mekong River discharge further increases and the monsoonal winds begin to change direction in September, the fresh anomalies begin propagating into the southwestern shallow shelf.

The seasonal mean and instantaneous snapshots of the modeled SSH in Figure 4-6, point also to the mesoscale structures that populate the SCS offshore the Vietnam coast. Both the WR and LR-WR simulations reproduces reasonably well shape and position of the *eddy pair*, an anticyclonic southern recirculation and a cyclonic northern recirculation that form every year at the end of spring off eastern Vietnam around 10-12°N, 110-114°E, which signature is apparent also in the temperature and salinity distributions. The anticyclonic recirculation has a warm and low salinity core, while the cyclonic one has a cold and salty core (Chen et al. 2010). The northern recirculation (called East Vietnam Eddy, hereafter EVE) is generally characterized by a narrow belt of low dynamic height east of Vietnam, is present in summer (June–August), and tends to strengthen toward the end of the season when prevailing winds are from the southwest (October–November). This cyclonic circulation persists in winter, until the monsoonal transition from the northeast to the southwest (March-May) (Qu et al., 2000). Over the summer, months long-lived anticyclonic eddies (up to three) form along the jet while it meanders bringing the river plume water offshore, extending west to east (see Figure 5 and 6 for the SSH evolution from June to August at the two resolutions). Cai and Gan (2017) used an idealized circular basin model to show that the anticyclonic eddy train originates from the separated jet and its downstream meandering, and that the stratification modulates intensity and vertical extension of the anticyclones. They analyzed the vorticity balance and found that its formation is due to baroclinic instability of the jet, as to be expected, as is favored by stratification.

## 5 Freshwater Dependence

### 5.1 Horizontal density gradient and frontal tendency

Freshwater anomalies are transported offshore to the deep portion of the basin by a joint effort of the VOC and mesoscale eddies. These anomalies in turn, modify the surface stratification. Additionally, near the Vietnam coast and along the jet, the riverine input enhances submesoscale activity and generates fronts that further promote freshwater advection (Luo et al., 2016). In the following, we consider the dynamical evolution of quantities such as density gradients and frontal tendency, comparing the runs with and without riverine forcing. We focus especially on the August to October period, when the freshwater impact is largest.

In our submesoscale permitting simulations, the seasonal evolution of lateral density gradients $|\nabla_h\rho|$ and frontal tendency $|FT|$ is exemplified in Figure 7 by snapshots and in Figure 8 by the time-series of the mean annual cycling for the WR and NoR runs. The flow frontal tendency, *FT*, defined as $FT = \frac{D|\nabla_h\rho|}{Dt} = Q \cdot \nabla_h\rho$ with $Q = (Q_1, Q_2) = -(\frac{du}{dx}\frac{d\rho}{dx} + \frac{dv}{dx}\frac{d\rho}{dy} + \frac{dv}{dy}\frac{d\rho}{dy})$ (Capet et al. 2008c; Hoskins 1982; Hoskins and Bretherton 1972) describes the evolution of the density gradients (i.e. their tendency to increase or to decrease). A positive frontal tendency indicates an increase of the magnitude of the density gradients over time and therefore frontogenesis, while a negative sign implies frontolysis. A frontal jet tends to be confluent and frontogenetic on the upstream side of a meander trough and diffluent and frontolytic on the downstream side of it (Bower et al., 1989; Thomas and Joyce 2010; Gula et al., 2016). In the WR case, surface density gradients and frontal tendency share a similar seasonal evolution, being greatest in September and October, and very small in spring. The seasonal cycle of $|\nabla_h\rho|$ and $|FT|$ follows that of the riverine input increasing from summer to late fall and decreasing from the end of December into spring, with a minimum in early February. $\nabla_h\rho$ and *FT* maps in NoR and WR in winter differ only near shore, around the Mekong rivermouth, given that the riverine input to the SCS is small and the monsoonal conditions favor downwelling. In summer, on the other hand, the much larger $\nabla_h\rho$ and F values in WR are controlled by the intrusion and offshore extension of the freshwater plume, dominating the contribution of coastal upwelling and mesoscale eddies. The WR extension covers a larger area offshore the shelf in the higher resolution case, and extends further to the north, through the advection of density anomalies into submesoscale fronts, which contribution is underestimated in the LR-WR case (Figure 7e and 7f).

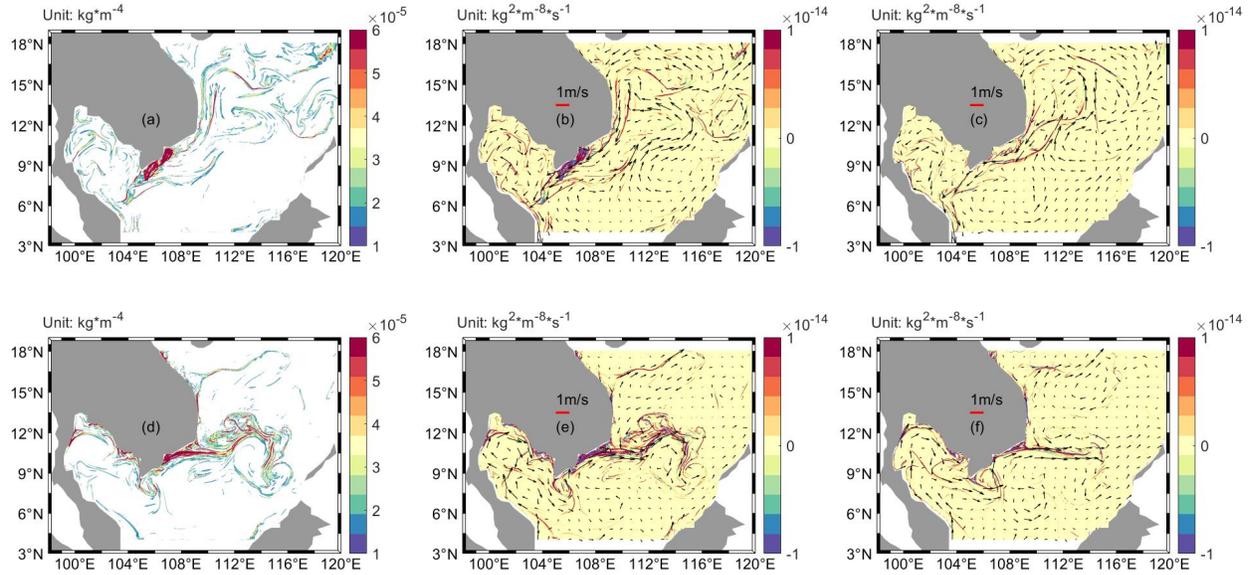

**Figure 7.** Instantaneous snapshots of horizontal density gradients (unit: kg m$^{-4}$) in the WR run in winter (12/15/2016) (a) and summer (08/05/2016) (d). Same days instantaneous snapshots of frontal tendency, unit: kg$^2$ m$^{-8}$ s$^{-1}$, in WR in (b - e) and NoR (c - f).

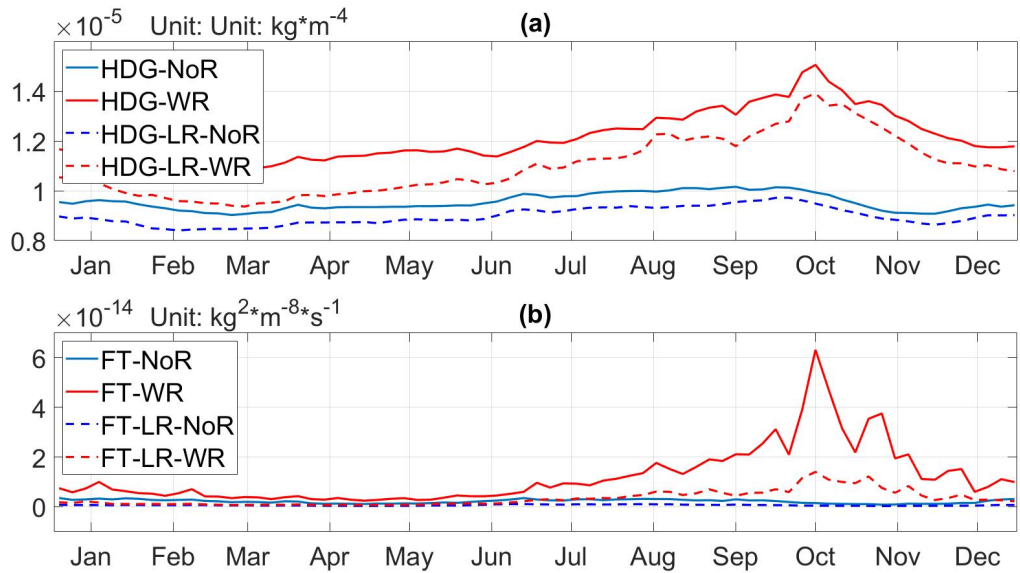

**Figure 8.** Annual cycle of (a) horizontal density gradients (HDG), unit: kg m$^{-4}$ and (b) absolute value of frontal tendency (FT), unit: kg$^2$ m$^{-8}$ s$^{-1}$ in the WR (red, solid), LR-WR (red, dashed), NoR (blue, solid) and NoR-WR (blue, dashed) simulations. FT time-series considering only positive values show comparable differences. All time-series are constructed using 5-day averages over the whole domain and integration period 2011-2016.

The presence/absence of water of riverine origin offshore influences the near-surface stratification and in turn the mesoscale circulation in the basin between, especially, August and October. Figure 9 shows the difference in sea surface heigh, horizontal velocity and stratification between the WR (LR-WR) and NoR (LR-NoR) simulations averaged over the six years considered and for August-to-October. Here the upper ocean stratification is defined by the density at 100 m minus the density at 5 m depth.

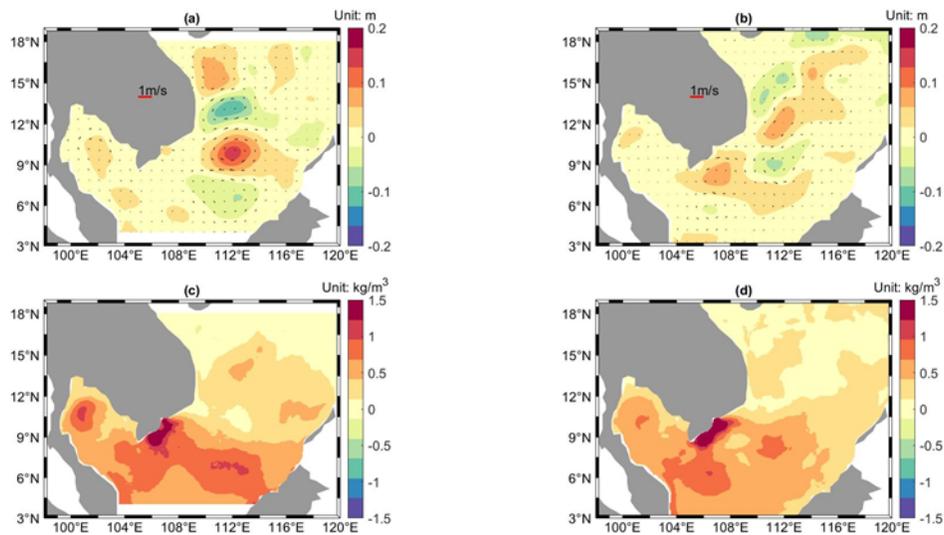

**Figure 9** Differences in the mean August-to-October SSH (a and b) and stratification (see text for definition) (c and d) fields averaged over 2011-2016 between runs with and without riverine input in the submesoscale permitting case (WR – NoR, a and c) and in the LR simulations (LR-WR – LR-NoR, b and d).

**5.2 Residual eddy train**

Independently of resolution and of the presence/absence of the Mekong freshwater fluxes, CROCO reproduces the eddy dipole structure and the EVE strengthening over the summer. The lack of accumulation of fresh and warm water near 10°N - 113°E in the NoR simulations, however, causes the offshore velocity to extend straightly eastward from 109°E to 112°E at approximately 9°N. The difference in SSH between WR and NoR runs has a noticeable wave (cyclonic-anticyclonic-cyclonic-anticyclonic) pattern east of the Vietnam coast with its strongest amplitude in September in all years, resulting in a greater northward component for the offshore velocity near 112°E. Among the residual mesoscale eddies, the clockwise residual eddy (hereafter indicated as CRE) centered around 10°N east off the coast is responsible for the stronger northward current around (11°N, 110°E) in WR. In the LR case the limited off-shore reach of the Mekong plume water implies smaller stratification differences away from the coast and a more coastally confined mesoscale residual.

Figure A1 shows the monthly time evolution from June to October of the modeled mean SSH and geostrophic velocities averaged over the 2011 - 2016 period in the WR solution and Figure A2 the difference (WR-NoR) for the same quantities. The typical pattern of the residual eddy train appears in June. From June to August, as the eddy dipole becomes more and more mature due to stronger SW monsoon and riverine input, the residual eddy train strengthens. This trend continues in September, despite a weakening of the SW monsoon winds, as the river discharge continues to increase, and the eddy train moves slightly to the south. With the onset of the NE monsoon in October and the reversal of the winds, the northern anticyclonic-cyclonic pattern weakens.

In agreement with the idealized simulations in Cai et al. (2017), the residual eddy train is linked to differences in the upper stratification in the SCS and its signature in the full SSH field is towards a strengthening of the anticyclones that develop in summer as shown in Figure 2. The change of size and magnitude, as well as the movement of the anticyclonic-cyclonic-anticyclonic pattern follows that of the riverine input which is transported according to the monsoon winds. Overall, the combined effect of topography and stratification drives the development of the residual eddies, among which the CRE is the major contributor (Figure 10).

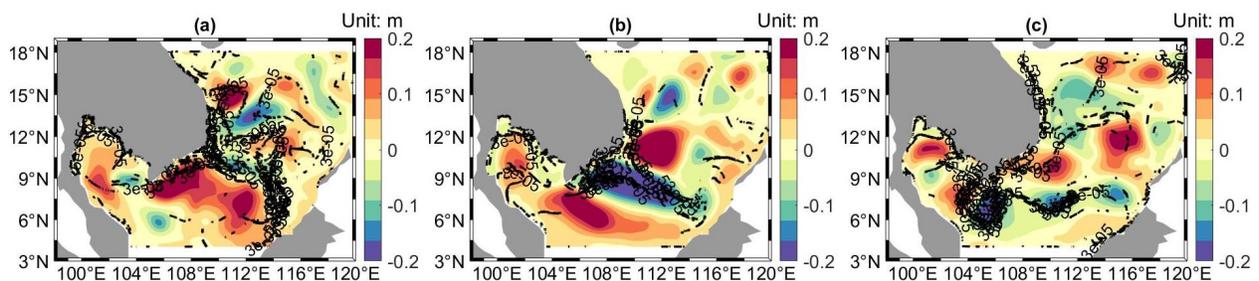

**Figure 10** Differences in the mean (a) August, (b) September and (c) October, 2014 SSH between WR and NoR with superposed the WR lateral density gradient contours on August 16, September

15 and October 15, 2014 comprised between $3 \times 10^{-5}$ and $5 \times 10^{-5}$ kg m$^{-4}$.

**5.3 Zonal and meridional velocity and volume transport**

Differences in mean lateral velocity averaged latitudinally over the area (9°N, 110°E) to (13°N, 115°E) from August to October, are shown in Figure 11 for the zonal and meridional components separately. The difference shows an enhancement of eastward momentum in WR. The largest difference, up to 0.15 ms$^{-1}$, is found in October, when the magnitude of the freshwater flux is strongest and the monsoon brings colder and saltier water from the Hainan province that pushes the fresh signals "back" into the area considered, followed by August, when the summer monsoon winds can push the near-surface water offshore more easily in presence of a shallower mixed-later in WR. The differences in the meridional component vary with longitude and throughout the season, with an overall stronger northward current in WR in each of the years considered and with a core extending to the base of the mixed-layer at about 40 m of depth. The magnitude of the meridional difference is largest in August.

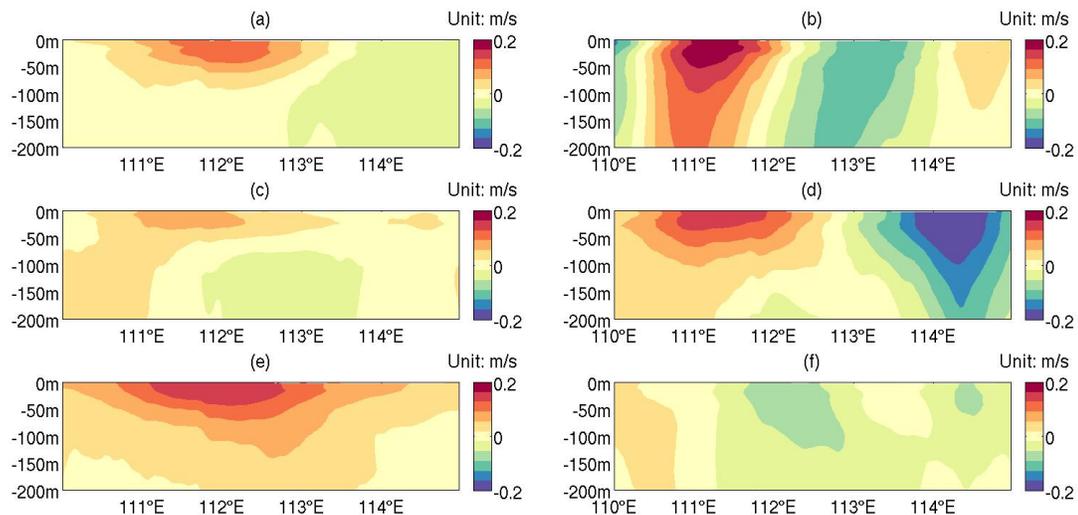

**Figure 11** Difference in zonal (left) and meridional (right) velocity components (WR – NoR) latitudinally averaged in the area (9°N, 110°E) to (13°N, 115°E) in August (a-b), September (c-d) and October (e-f). Unit: ms$^{-1}$.

Finally, we computed the west-to-east (WE) and south-to-north (SN) volume transport at 12°N, integrating the velocities at the given section following Sanchez-Roman et al. (2018) (Figure 12). In the presence of riverine input, the WE volume transport in the transect is predominantly eastward in the mixed-layer in WR.

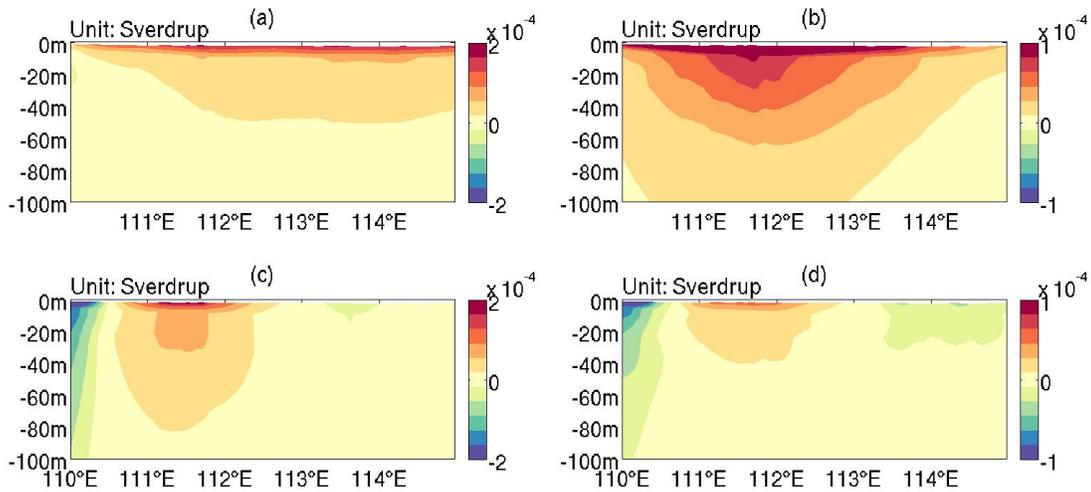

**Figure 12** West-to-east (a-b) and south-to-north (c-d) volume transport across a transect at 12°N in WR, and the WR-NoR difference averaged over August-to-October. Unit: Sverdrup.

The warm and fresh water brought from the coast to 12°N, leads to higher SSH and stronger divergence, and an increased eastward volume transport in the presence of riverine forcing. In the meridional direction, the volume transport in WR has a stronger northward component, especially in August and September, close to the surface.

**6 Conclusions**

Through a modeling exercise we have quantified the dynamical impact of Mekong River plume on the SCS circulation during the summer and early fall seasons. We have shown that the changes in stratification caused by the freshwater fluxes are responsible for a residual eddy train. The residual mesoscale circulation emerges in June and strengthens until August, disappearing with the monsoon wind reversal by the end of October.

A comparable impact of freshwater inputs on the overall circulation was found in the Caspian Sea (Kara et al., 2010). In that case, the simulation performed with no river discharge still produced the observed seasonal variation in surface circulation in the Caspian Sea, but resulted in a decrease in intensity of the southward current along the western coast of the central Caspian Sea, impacted by the Volga River. In the SCS, on the other hand, the summer 'gyres' and coastal currents are wind-driven and alongshore coastal currents are not significantly modified. However, the freshwater-induced mesoscale anomalies introduce a northward shift in the surface jet that forms offshore the Vietnam coast. The jet meanders and is more prone to baroclinic instability due to the increased stratification whenever influenced by the riverine input, in agreement with Cai and Gan's idealized work (2017). Such northward component in the jet velocity is indeed observed in satellite SSH maps, and is greatly underestimated if the river fluxes are neglected.

Furthermore, we explored the role of submesoscale circulations and more broadly model resolution in representing the dynamical feedback of the riverine input in the SCS. By comparing mesoscale resolving runs at 5 km horizontal resolution with submesoscale permitting simulations at 1.6 km horizontal resolution, we found that submesoscale fronts contribute to the offshore transport of freshwater anomalies and, in doing so, to the stratification patterns. The end result is an eastward, offshore shift of the residual eddy train in the solution at higher resolution.

Overall, our results point to the need to include ocean circulation changes - and in turn primary productivity changes not just over the shelf and in the alluvial plains of Cambodia and Vietnam (Yoshida et al., 2020), but also in the open ocean, being the Mekong River a major supplier of nitrogen and phosphorus among other nutrients (Weber et al., 2019) - to the list of environmental impacts associated with the construction of over 100 hydropower dams along the Mekong River Basin and its major tributaries.

We note that the investigation performed, while realistic, did not include tides, which may enhance mixing in this basin, and likely underestimated the Mekong plume impact by imposing the freshwater fluxes as salinity anomalies, therefore neglecting their contribution to the momentum flux. We plan to further explore those aspects in the near future.

**Acknowledgments, Samples, and Data**

This research has been supported by the National Science Foundation (grant OCE-1658174). The authors wish to thank Dr. Patrick Marchesiello, who provided the Mekong River data. We also thank the Schmidt Ocean Institute for supporting a 2016 cruise off the coast of Vietnam, during which the idea behind this work formed.

All model runs used in this research are made available at the dropbox address: https://www.dropbox.com/sh/26qudw20m99nisy/AAA7Yq4hZ5bBE65072sjtAy-a?dl=0 . The Georgia Institute of Technology has a campus-wide license for Dropbox, providing unlimited, ITAR-compliant cloud storage space for large modeling data-sets.

Model configuration files, and MATLAB codes for all analysis are publicly available through fig**share** (DOI: https://doi.org/10.6084/m9.figshare.14589165.v5).

CROCO is a community model. The version used in this work is 1.0 and can be found at https://www.croco-ocean.org/download/croco-project/.

The ERA5 data-set used to force CROCO was downloaded from: Copernicus Climate Change Service (C3S) (2017): ERA5: Fifth generation of ECMWF atmospheric reanalyses of the global climate, Copernicus Climate Change Service Climate Data Store (CDS) https://cds.climate.copernicus.eu/cdsapp#!/home. Last date of access: May, 11 2021

The SODA 3.4.2 reanalysis data were downloaded from: https://www2.atmos.umd.edu/~ocean/index_files/soda3.4.2_mn_download_b.htm. Last date of access: May 11, 2021

**Appendix**

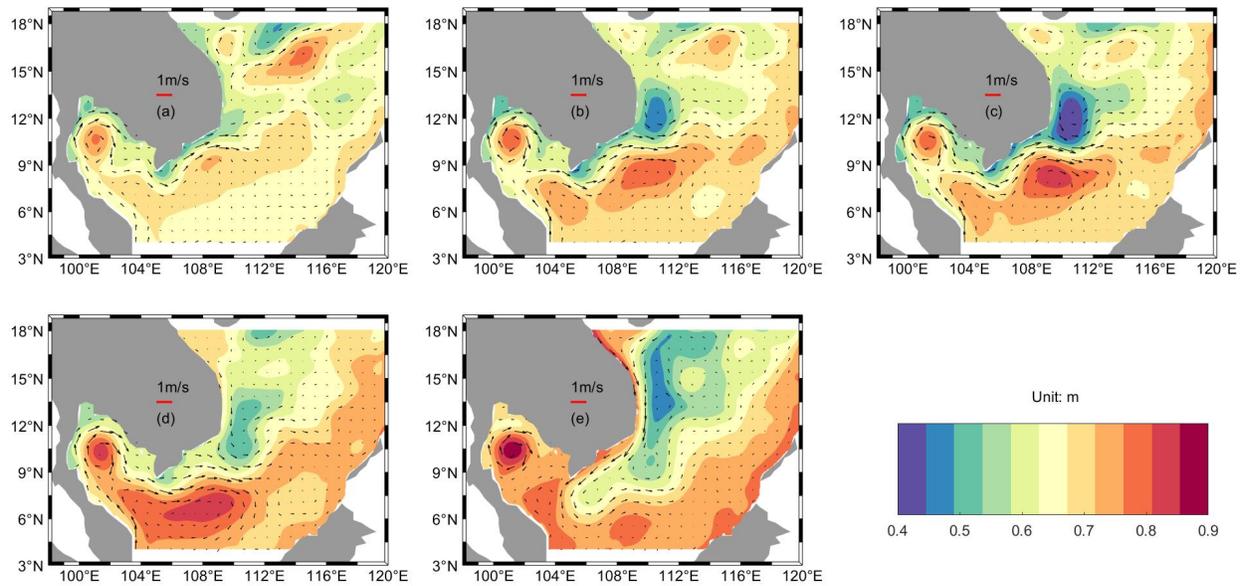

**Figure A1** Time evolution of monthly mean modeled SSH with superposed geostrophic velocities averaged over the 2011-2016 period in June (a), July (b), August (c), September (d) and October (e) in the WR simulation.

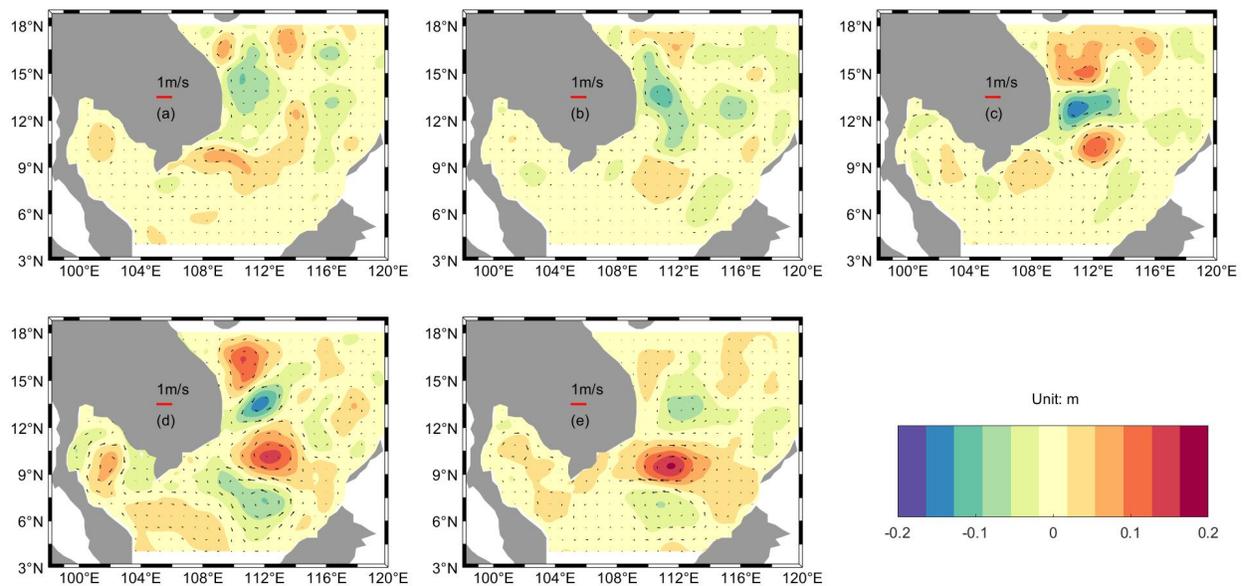

**Figure A2** As in figure A1 but for the (WR-NoR) difference in monthly SSH and geostrophic velocity, June to October.